\documentclass[aps,prl,twocolumn,floatfix,preprintnumbers,showpacs]{revtex4}
\usepackage{graphicx}
\usepackage{amsmath}
\usepackage{amsfonts}
\usepackage{color}
\usepackage[dvips]{draftcopy}

\newcommand{\etal}{{\em et al.}}

\newcommand{\threejsymbol}[6]{\left(\begin{array}{ccc} #1 & #2 & #3 \\ #4 & #5 & #6 \end{array}\right)}            
\begin{document}

\preprint{\rm LLNL-JRNL-409701}

\title{
       Efficient and robust calculation of femtoscopic correlation functions 
       in spherical harmonics directly from the raw pairs measured in heavy-ion
       collisions
}
\author{A.~Kisiel} 
\affiliation{Department of Physics, The Ohio State University, Columbus OH 43210, USA}

\author{D.A.~Brown} 
\affiliation{Lawrence Livermore National Laboratory, Livermore California 94551}

\date{\today}

\begin{abstract} 
We present the formalism for calculating the femtoscopic correlation
function directly in spherical harmonics. The numerator and
denominator are stored as a set of one-dimensional histograms
representing the spherical harmonic decompositions of each. We present 
the formalism to calculate the correlation function from them directly,
without going to any three-dimensional histogram.  We discuss the
practical implementation of the method and we provide an example
of its use.  We also discuss the stability of the method in the presence of
angular holes in the underlying data (e.g. from experimental
acceptance). 
\end{abstract}

\pacs{PACS numbers: 25.75.-q, 25.75.Gz}

\maketitle

Spherical harmonics are one of the most commonly used mathematical
tools for the analysis of experimental data.  For example,
geopotential models of the Earth's gravitational field are matched to
experimental data up to harmonic order $\ell_{max} = 70$ \cite{GGM02}.
Measurements of the cosmic microwave background is expanded to by the
Cosmic Background Explorer (COBE) \cite{COBE} and successor
experiments extend further to an impressive $\ell_{max}\approx 1500$
\cite{WMAP}.  In the femtoscopic measurements of particle emitting
sources in heavy-ion collisions, one also finds it useful to expand the
measured correlations and extracted sources in spherical harmonics
\cite{Brown05,Chajecki:2008vg}. 

In nearly all applications of spherical harmonics to data analysis and
reduction, one is faced with the problem of ``holes in the data'' --
i.e. sampling bias \cite{Tenorio}.  In the COBE analysis mentioned
above, this bias occurs because the Milky Way masks a sizable solid
angle of the sky.  When constructing potential maps of the earth, the
problem is even more severe as one uses strips of data obtained from
various satellite and balloon-borne experiments to derive the map
\cite{Blais02}. In heavy-ion collisions, the sampling bias most often
arises because the detector acceptance does not span all of
phase-space.  

Unlike other applications, the femtoscopic correlation functions are
actually ratios of two single particle distributions: the true pair
distribution (the numerator) and the mixed pair distribution (the
denominator).  Thus, expanding the correlation function in spherical
harmonics is a more involved than expanding the underlying single
particle distributions. It is often impractical to simply bin the two singles
distributions in 3D histograms and make a ratio since one does
not have a meaningful number of pairs in each bin, either due to the
statistics of particle production or due to detector acceptances.
Furthermore, since the end result is a spherical harmonic
decomposition of the correlation, one must have absurdly high
harmonics in one's decomposition to resolve high-momentum bins which
have small angular extent and marginal statistics. 

Rather than pursue this, we adopt an alternate approach: we construct the
raw pair distributions directly in spherical harmonics, then we extract
the correlation function, also in spherical harmonics, by viewing the ratio
as an inverse problem.  A feature of this approach is that one
preserves the full cross-$\ell,m$ data covariance that is currently
ignored when imaging 3D correlations \cite{Brown05}.     

We now outline this paper.  In the first section, we detail how we expand the
numerator and denominator distributions in spherical harmonics and how we pack 
them into the vectors and matrices for further manipulation.  In the second 
section, we describe how to compute the correlation in spherical harmonics using 
these harmonic expansions.  In the last section, we demonstrate the technique in 
realistic examples.  Further, we show how this technique is insensitive to the 
sampling bias imposed by sizable holes in pair acceptance.  

%\debug{
%From a mathematical point of view, to perfectly represent any 3D
%function in spherical harmonics one needs an infinite number of
%$Y^{\ell}_{m}$ components. The reason why the decomposition is useful in
%femtoscopy is that in the specific case of a femtoscopic correlation
%function only very few of these components are needed (a practical
%limit is $\ell \leq 4$). All the others are either mathematically
%required to vanish or do not carry relevant physics information. In
%other words the spherical harmonics is a particularly efficient
%representation: it extracts, from the 3D correlation function, only
%the relevant information, and stores it as few 1D histograms. This
%allows for important simplification of analysis methods and 
%significant reduction of computational resources consumption. The
%decomposition methods used up to now, performed this extraction at the
%correlation function level. However the raw pair distributions: the
%``signal'' and the ``background'', were still constructed in an 
%inefficient way: as 3D histograms. In this work we present the
%decomposition method which performs the extraction of the relevant
%information already at the level of filling the raw distributions. As
%a result the efficient representation is used at all stages of the
%analysis, there is no need to use 3D histograms. The price to pay for
%this improvement is the increased computational time for data
%analysis: one needs to calculate the spherical harmonics weight for
%each pair.
%}

\section{Constructing the numerator and denominator}

The femtoscopic correlation function, $C({\bf q})$, is defined as a
ratio of the probability to observe a correlated pair of particles
at a given relative momentum in the same event (numerator), $T({\bf
q})$, to the probability to observe such a pair in an uncorrelated
state (denominator), $M({\bf q})$: 
\begin{equation}
T({\bf q}) = C({\bf q}) M({\bf q}).
\end{equation}
The uncorrelated distribution is usually obtained by mixing particles
from different events. Here the relative momentum is given in the pair
center of mass frame as ${\bf q}=\frac{1}{2}({\bf p_1}-{\bf p_2}) =
{\bf k}^*$ for particles 1 and 2.  Here ${\bf k}^*$ is the  notation
commonly used for femtoscopic correlations for pairs of two different
types of particles, and will be used interchangeably with ${\bf q}$
(traditionally used in femtoscopic correlations of pairs of identical
partilces) later in the paper. We expand the numerator, denominator
and correlation function in spherical harmonics, e.g.: 
\begin{equation}
 T({\bf q}) = \sqrt{4\pi}\sum_{\ell m} T_{\ell m}(q) Y_{\ell m}(\Omega_{\hat{\bf q}}). 
\end{equation}

\begin{table}
\begin{ruledtabular}
\begin{tabular}{lll}
& {\bf Condition}& {\bf Relation} \\\hline\hline
1 & Distribution is real &  $T_{\ell m} = T_{\ell -m}^*$ \\
2 & $x \rightarrow -x$ symmetry & $T_{\ell m}(x,y,z) = (-1)^m T_{\ell m}^*(-x,y,z)$ \\
3 & $y \rightarrow -y$ symmetry & $T_{\ell m}(x,y,z) = T_{\ell m}^*(x,-y,z)$ \\
4 & $z \rightarrow -z$ symmetry & $T_{\ell m}(x,y,z) = (-1)^{\ell+m} T_{\ell m}(x,y,-z)$ \\
5 & ${\bf r} \rightarrow -{\bf r}$ symmetry & $T_{\ell m}(x,y,z) = (-1)^\ell T_{\ell m}(-x,-y,-z)$ 
\end{tabular}
\end{ruledtabular}
\caption{Conditions on the distribution imply relations between the
  different terms in the $Y_{\ell m}$ expansion.  Condition 1 is
  always valid for the true and mixed pair distributions and the
  correlation function.  Condition 5 is always valid for like pair
  correlation and the corresponding pairs distributions.  By
  exploiting symmetries in 2-5, we can reduce the number of components
  in our data vectors.
  \label{table1}}
\end{table}

We can compute the pairs distributions directly in spherical harmonics
by observing that 
\begin{equation}
 T_{\ell m}(q)  = \frac{1}{\sqrt{4\pi}} \int_{4\pi} d\Omega_{\hat{\bf q}} T({\bf q})Y^{*}_{\ell m}(\Omega_{\hat{\bf q}}). 
\end{equation}
can be built up by summing over the pairs, which is essentially a Monte-Carlo integration process: 
\begin{equation}
 T_{\ell m}(q_{n})  \approx  \frac{\sqrt{4\pi}}{N}  \sum_{i=1}^{N}\left\{ \begin{array}{rl}   Y^{*}_{\ell m}(\Omega_{\hat{\bf
 q}_{i}} ) & \textrm{if}\;q_{i}\;\textrm{in bin}\;n,  \\
 0 & \textrm{otherwise.} \end{array}      \right.
\label{eq:tfrompairs}
\end{equation}   
Here, $N$ is the number of pairs in the spectrum.
In our approach a pair is added to $T_{\ell m}(q_{n})$ ($M_{\ell m}(q_{n})$) in the following way. 
First we calculate the relative momentum $\bf q$ and decompose it into 
$|{\bf q}|$ (which determines the 1D $q$-bin number), $\theta_{\hat{\bf q}_{i}}$ and
$\phi_{\hat{\bf q}_{i}}$. Having the angles, we can calculate the spherical
harmonics functions, usually up to some limiting value of $\ell$. Then
the pair is added to all histograms of the corresponding function,
with weights equal to the respective $Y_{\ell m}$'s.  We could eliminate 
many of the components of $T_{\ell m}(q_{n})$ and $M_{\ell m}(q_{n})$ by taking advantage of 
the symmetries in Table I, but we have chosen to keep all components so that
we can perform cross-checks of our work.  This does introduce complications when
performing some matrix manipulations, as we discuss in the following section.

The covariance can be built in a similiar way by noting 
($\textrm{if}\;q_{i}\;\textrm{in radial bin}\;n$ only):   
\begin{equation}
\begin{split}
    \lefteqn{ \Delta^{2} T_{\ell m\ell' m'}(q_{n})  \approx \frac{4\pi}{N(N-1)} \times }&\\
    & \sum_{i=1}^{N}  \left( Y^{*}_{\ell m}(\Omega_{\hat{\bf q}_{i}}) - \frac{T_{\ell
    m}(q_{n})}{\sqrt{4\pi}}\right)
    \left( Y^{*}_{\ell' m'}(\Omega_{\hat{\bf q}_{i}}) - \frac{T_{\ell' m'}(q_{n})}{\sqrt{4\pi}}\right)^*.
\end{split}    
\end{equation}   
Taking the diagonal elements (i.e. $\ell m = \ell' m'$), we find (uncorrelated) 
uncertainties of    
\begin{equation}
 \Delta T_{\ell m}(q_{n})  \approx  \sqrt{\Delta^{2} T_{\ell m\ell m}(q_{n})}.
\end{equation}   

In contrast with our approach, in the traditional representation both 
the numerator and denominator were
stored as 3D histograms, using either the Bertsch-Pratt
coordinates \cite{BPCoords} in Cartesian form $q_{out}$, $q_{side}$ and $q_{long}$, or
in spherical: $|{\bf q}|$, $\cos(\theta_{\bf \hat{q}})$, $\phi_{\bf \hat{q}}$. 
In the traditional representation the numerator (denominator) is a 3D
histogram, and each signal (background) pair is added to exactly one
bin of this histogram with weight 1.0.  This representation has several
disadvantages: one needs significant statistics to have a
meaningful number of pairs in each bin, single-particle momentum
acceptance can result in ``holes'' or empty bins in two-particle, or
relative momentum space, and the binning corrections need to be
applied. Also going to higher moments in the decomposition
requires larger number of bins on the $\phi_{\bf \hat{q}}$ and $\theta_{\bf \hat{q}}$ direction.

In practice, we store both the real and imaginary parts of the 
numerator and the denominator as an array of
one-dimensional histograms in $q = |{\bf q}|$, as seen in
Eq.~\eqref{trepresntation}, which is represented as the vector ${\bf T}_q$: 
\begin{equation}    
{\bf T}_q = \left( 
\begin{array}{c}
T_{00}(q)\\
T_{10}(q)\\
\Re T_{11}(q)\\
\Im T_{11}(q)\\        
T_{20}(q)\\    
\Re T_{21}(q)\\
\Im T_{21}(q) \\
\Re T_{22}(q)\\
\Im T_{22}(q) \\
T_{30}(q)\\    
\vdots            
\end{array}    
\right)            
\label{trepresntation}
\end{equation}
There is also a corresponding covariance matrix ${\bf \Delta}^{2}{\bf T}_q$,
which is a two-dimensional matrix for each $q$-bin and is packed in an analogous fashion.

\section{Calculating the correlation function}

Since we have expanded the pair distributions and the correlation in spherical 
harmonics, we have
\begin{equation}
\begin{split}
T_{\ell m}(q) = & \sum_{\ell' m'\ell'' m''}
%     {
%	\begin{array}{c}
%		_{\ell' m'}\\
%		_{\ell'' m''}
%	\end{array}
%	} 
	M_{\ell' m'}(q) C_{\ell'' m''}(q) \\
    & \times \int_{4\pi} d\Omega_{\hat{\bf q}}Y_{\ell m}^{*}(\Omega_{\hat{\bf q}}) Y_{\ell' m'}(\Omega_{\hat{\bf q}}) Y_{\ell'' m''}(\Omega_{\hat{\bf q}}) \\
    \equiv & \sum_{\ell'' m''} \tilde{M}_{\ell m\ell'' m''}(q) C_{\ell'' m''}(q). \label{eq:getClmDirectly}    
\end{split}    
\end{equation}    
With the packing in Eq. \eqref{trepresntation}, equation \eqref{eq:getClmDirectly} can be written very compactly: ${\bf T}_q = {\bf \tilde{M}}_q\cdot{\bf C}_q$.  Eq. \eqref{eq:getClmDirectly} gives us a way to compute $C_{\ell m}(q)$ directly from the pair distributions expanded in spherical harmonics.  
\begin{widetext}
Here the ${\bf \tilde M}_q$ matrix is written in terms of Wigner 3-$j$ symbols as:
\begin{equation}
\tilde M_{\ell m \ell''m''} = \sum_{\ell'm'}M_{\ell'm'}(q) (-1)^{m}
\sqrt{(2\ell+1)(2\ell'+1)(2\ell''+1)} \threejsymbol{\ell}{\ell'}{\ell''}{0}{0}{0}
\threejsymbol{\ell}{\ell'}{\ell''}{-m}{m'}{m''}. 
\label{eq:mtilde}
\end{equation} 
\end{widetext}

%Now, how do we extract the correlation terms using this result?  As in
%all things, there is a wrong way and a right way.    We'll start with
%the wrong way.  Since ${\bf T}$ and ${\bf C}$ have the same
%dimensions, we could just invert the $\tilde{M}$ matrix, giving: 
%\begin{equation}                
%{\bf C} = \tilde{M}^{-1}\cdot{\bf T}.
%\end{equation}    
%Propagating the uncertainties, we find
%\begin{equation}                
%\Delta^{2}C = \tilde{M}^{-1}\cdot\Delta^{2}T\cdot(\tilde{M}^{-1})^{T}.                               
%\end{equation}
%On the face of it, that looks OK, so what is wrong with it?  Well, because we did no weighting with respect to the uncertainties in the pair distributions, we've given equal weight to both the good and bad data.  For example, near $q=0$, $T({\bf q}), M({\bf q})$ and $C({\bf q})$ must be nearly isotropic.  This means that for example $T_{\ell m}(q)$ for $\ell m \ne 00$ should be small, but with possibly large uncertainties.  This would then propagate into the correlation terms in possibly bad ways.  When we do the correlation extraction this way, we are really minimizing this quantity:
%\begin{equation}                
%|{\bf T} - \tilde{M}\cdot{\bf C}|^{2}    
%\end{equation}
    
In order to calculate the correlation function, we view the problem as an inverse problem.  
To solve it, one needs to minimize the $\chi^{2}$: 
\begin{equation}                
({\bf T}_q - {\bf \tilde{M}}_q\cdot{\bf C}_q)^{T}\cdot({\bf \Delta}^{2}{\bf T}_q)^{-1}\cdot({\bf T}_q - {\bf \tilde{M}}_q\cdot{\bf C}_q).
\end{equation}
The formula uses the full covariance matrix in the true distribution, but not in the 
mixed pair distribution.  Because the mixed pair distribution is constructed by 
pairs from different events, it is not limited by statistics and can be computed to
arbitrarily high precision, making the uncertainties negligible for our purposes.
The problem of minimizing the $\chi^{2}$ is identical to the one posed by the imaging
procedure in Ref. \cite{Brown05} and the solution is well known:  
\begin{equation}                
{\bf C}_q = {\bf \Delta}^{2}{\bf C}_q\cdot{\bf \tilde{M}}_q^{T}\cdot({\bf \Delta}^{2}{\bf T}_q)^{-1}\cdot{\bf T}_q.
\label{cformula}
\end{equation}
where the covariance is also calculated:
\begin{equation}
{\bf \Delta}^{2}{\bf C}_q=({\bf \tilde{M}}_q^{T}\cdot({\bf \Delta}^{2}{\bf T}_q)^{-1}\cdot{\bf \tilde{M}}_q)^{-1}.
\label{covformula}
\end{equation}
The uncorrelated uncertainties of the correlation are just the square root of the trace of the covariance ${\bf \Delta}^{2}{\bf C}_q$: ${\bf \Delta}{\bf C}_q = \sqrt{ Tr {{\bf \Delta}^{2}{\bf C}_q }}$.  Written out, we are simply taking the diagonal elements (i.e. $\ell m = \ell' m'$) and finding (uncorrelated) 
uncertainties of    
\begin{equation}
 \Delta C_{\ell m}(q_{n})  \approx  \sqrt{\Delta^{2} C_{\ell m\ell m}(q_{n})}.
\end{equation}   
With the error propagation done in this way, one gets the cross-$\ell m$
correlations in the covariance matrix by virtue of the cross-$\ell m$
correlations built into the ${\bf \tilde{M}}_q$ and ${\bf \Delta}^{2}{\bf T}_q$ matrices. 

We compute the correlation function according to
Eq.~\eqref{cformula}. As this involves several matrix inversions, it is
important to make sure that the matrix determinant is not zero. That
means that one cannot use $\ell m$ combinations that are a linear
combination of other $\ell m$'s. As all of the pair distributions are real, 
we cannot keep $m>0$ and $m<0$ components at the same time 
(see Table \ref{table1}).  Therefore we
adopt the convention to only use positive components (including $m=0$ 
component).  Therefore we remove all
functions with negative $m$ from both ${\bf T}_q$ and ${\bf M}_q$ when solving
Eq.~(\ref{cformula}).  One can add the missing
negative $m$ components to the correlation function by multiplying
the positive $m$ values by the appropriate factor of
$(-1)^{\ell+m}$. For consistency one may repeat the procedure, this time
removing the positive $m$ components from ${\bf T}_q$ and ${\bf M}_q$ and
the obtained results should be identical. 

We remind the reader that each $q$ bin is independent in ${\bf T}_q$, ${\bf M}_q$ and
${\bf C}_q$. Therefore solving Eq.~\eqref{cformula} can be done for each $q$ bin
independently. The starting points then are not vectors of functions,
but simply vectors of real numbers ${\bf T}_{q}$ and ${\bf M}_{q}$ and the result
is also a vector of real numbers ${\bf C}_{q}$.  A covariance matrix $({\bf \Delta}^{2}{\bf
T})_{q}$ is in this case a 2D matrix of real numbers, as is the
resulting correlation function covariance matrix $({\bf \Delta}^{2}{\bf C})_{q}$. Solution of
Eq.~\eqref{cformula} then reduces to a problem of solving a set of linear
equations, for which many standard numerical algorithms exist. The
procedure is repeated for each $q$ bin and the ${\bf C}_q$ vector is
filled in steps.

\section{Realistic Examples}

We have tested and applied the formalism to the construction of
two-particle correlation functions for identical and non-identical particles in
relativistic heavy-ion collisions. Our tests include a study of the robustness
of our approach in the presence of $\theta-\phi$ acceptance holes in relative 
momentum, a common occurance in experiments including the STAR experiment of which 
one of the authors (Kisiel) is a collaboration member.  
Below we present some tests of the method with a realistic model. 
%As we work with non-identical correlations, we switch notation 
%for the relative momentum from $q$ to the commonly used $k^*$.

\begin{center}
\begin{figure*}[tb]
\includegraphics[angle=0,width=0.9 \textwidth]{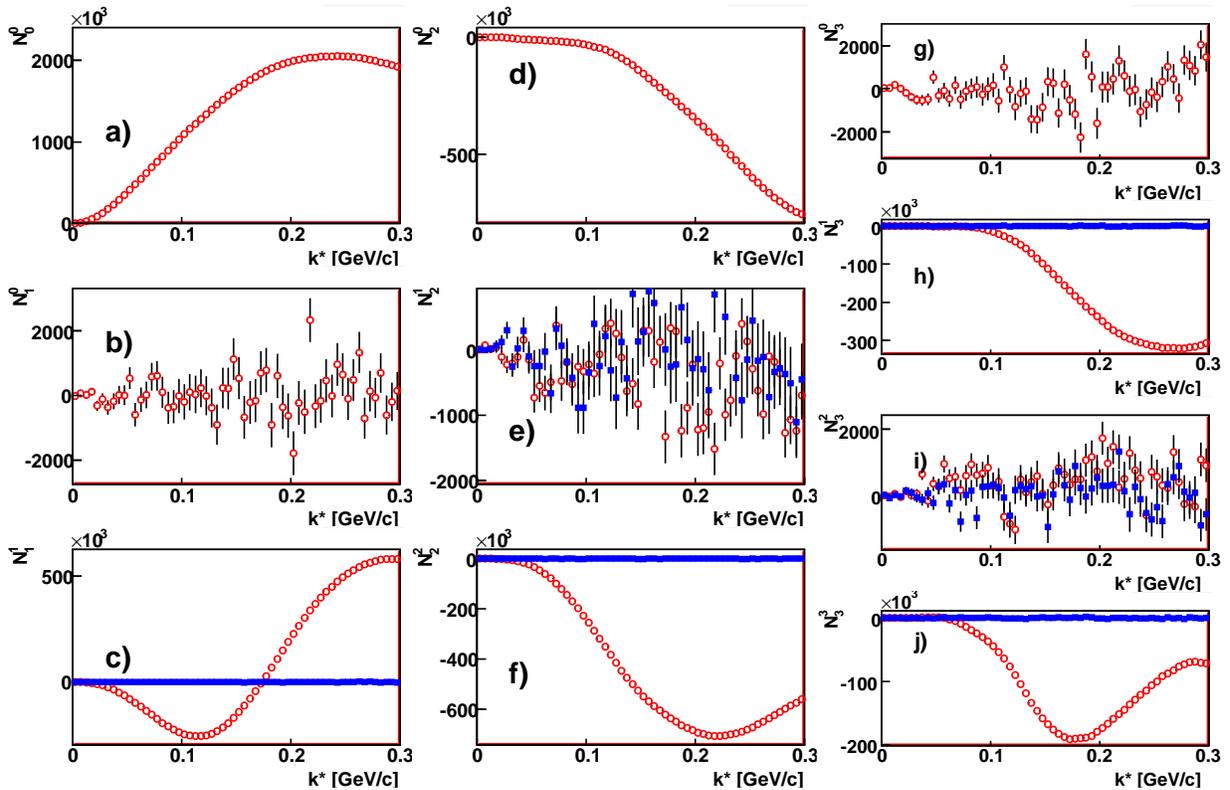}
%\vspace{-5mm}
\caption{(Color online) Numerator of an example $\pi^+K^+$ correlation
  function binned directly in spherical harmonics, as a function of
  the first particle's momentum in the pair rest frame ${\bf
    k}^{*}$. Panel a) shows $\ell=0$, panels b) and c) show $\ell=1$
  components, center panels d), e) an f) shows $\ell=2$ components,
  right panels g), h), i) and j) show $\ell=3$ components. Red open circles
  represent real part of the decomposition, blue closed triangles show
  imaginary part.} 
\label{fig:numlm}
\end{figure*}
\end{center}

\subsection{Example correlation functions}

Two particles are femtoscopically correlated if they have a small
relative momentum ${\bf k}^{*}$ in the pairs' rest frame (for identical
particles we use ${\bf q}={\bf k}^{*}$). If they are not identical and have
different masses, they must have different momenta in the laboratory
frame. In such case it can happen that due to specific momentum
acceptance of the experiment, pairs with specific values of $k^*=|{\bf k}^{*}|$
and certain combinations of polar and azimuthal components of ${\bf 
k}^{*}$ cannot be measured.  In terms of the spherical harmonic
representation, this results in a hole in the pair acceptance for
certain regions in $k^*$, $\phi$ and $\cos \theta$. This is observed
e.g. for pion-kaon pairs in the STAR experiment. Such a hole
presents a methodological problem for traditional methods of
decomposing the correlation function in spherical harmonics as
they rely on the existence of certain symmetries in pair
distributions.  In particular, they assume that the multiplicity of
pairs with a given $k^*_{out}$ is equal to the multiplicity of pairs
with $-k^*_{out}$. For non-identical particles there is no such
symmetry.  While it is certainly possible to improve the existing
methods and to remove this dependence, we propose to move to the more
advanced decomposition method presented in this paper and bypass the
problem altogether. In our method, this hole is reflected in both the
numerator and denominator by a lower number of pairs contributing at
some bins of $k^*$.

Examples of the numerator of the correlation function, binned
directly in spherical harmonics, are shown in Fig.~\ref{fig:numlm}. In
this case, the distributions result from a simulation of the $\pi^+K^+$ correlation 
function in the Therminator model using the STAR detector acceptance 
\cite{Kisiel07}. The acceptance is symmetric with respect to $\cos{\theta}$
so the $\ell,m=1,0$ component vanishes~\cite{Chajecki:2008vg}. All the imaginary 
components vanish. Also the $\ell,m=2,1$ as well as $\ell,m=3,0$ and $\ell,m=3,2$ 
vanish due to polar angle symmetry.  Apart from that, the numerator shows
non-trivial structure both as a function of $k^*$ and $\phi$.  The
synergy between spherical harmonic decomposition and femtoscopic
correlation function is nicely illustrated in this plot. A full 3D
information, which in traditional 3D implementation would require tens
of thousands of bins to store, is reduced to a few 1D histograms. Out
of these only a select few carry important information, while others
conveniently vanish due to the intrinsic symmetries of the pair
distribution. The significance of the components diminishes with growing
$\ell$, ensuring that cutting the decomposition at some $\ell_{max}$
should not distort the function. 

\begin{center}
\begin{figure*}[tb]
\includegraphics[angle=0,width=0.9 \textwidth]{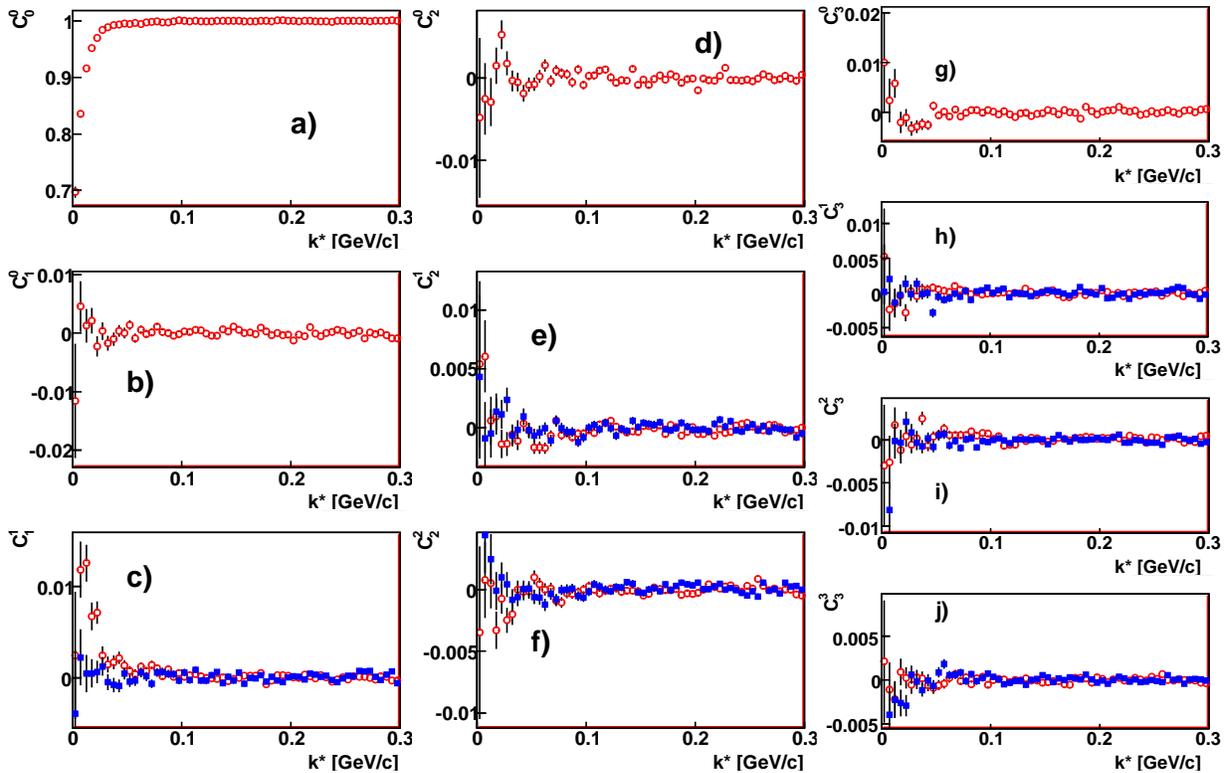}
%\vspace{-5mm}
\caption{(Color online) The example correlation function binned
directly in spherical harmonics.  The red open circles are the real parts of 
each term and the filled blue squares are the imaginary part of each term.
The spherical moments of the numerator spectra 
are shown on panels as in Fig.~\ref{fig:numlm}. }
\label{fig:cfnlm}
\end{figure*}
\end{center}

Having in mind that the underlying numerator has a non-trivial
structure both in $k^*$ and $\phi$ it is interesting to see how the
correlation function itself, calculated with the method above,
behaves. It is shown in Fig.~\ref{fig:cfnlm}. Again, the imaginary
components all vanish, as they should. The $C_{00}$ component shows
the expected behavior coming from a Coulomb repulsion of same charge
pion and kaon. The $C_{20}$ and $C_{22}$ components show small
deviations from zero, which signals the fact that the size of the
underlying system is not the same in the $out$, $side$ and $long$
directions. Also the $C_{11}$ component deviates from zero
significantly, a signature of the average emission point asymmetry
between pions and kaons. In summary the example confirms several
important points: (a) the correlation function can be calculated via
the direct $Y_{\ell m}$ method. (b) The important physics signals in
$C_{00}$, $C_{11}$, $C_{20}$ and $C_{22}$, are
preserved. (c) Other components of the correlation function vanish, as
they should - an important cross-check of the method.

\subsection{Limits of applicability in presence of an acceptance hole}

\begin{center}
\begin{figure*}[tb]
\includegraphics[angle=0,width=0.75 \textwidth]{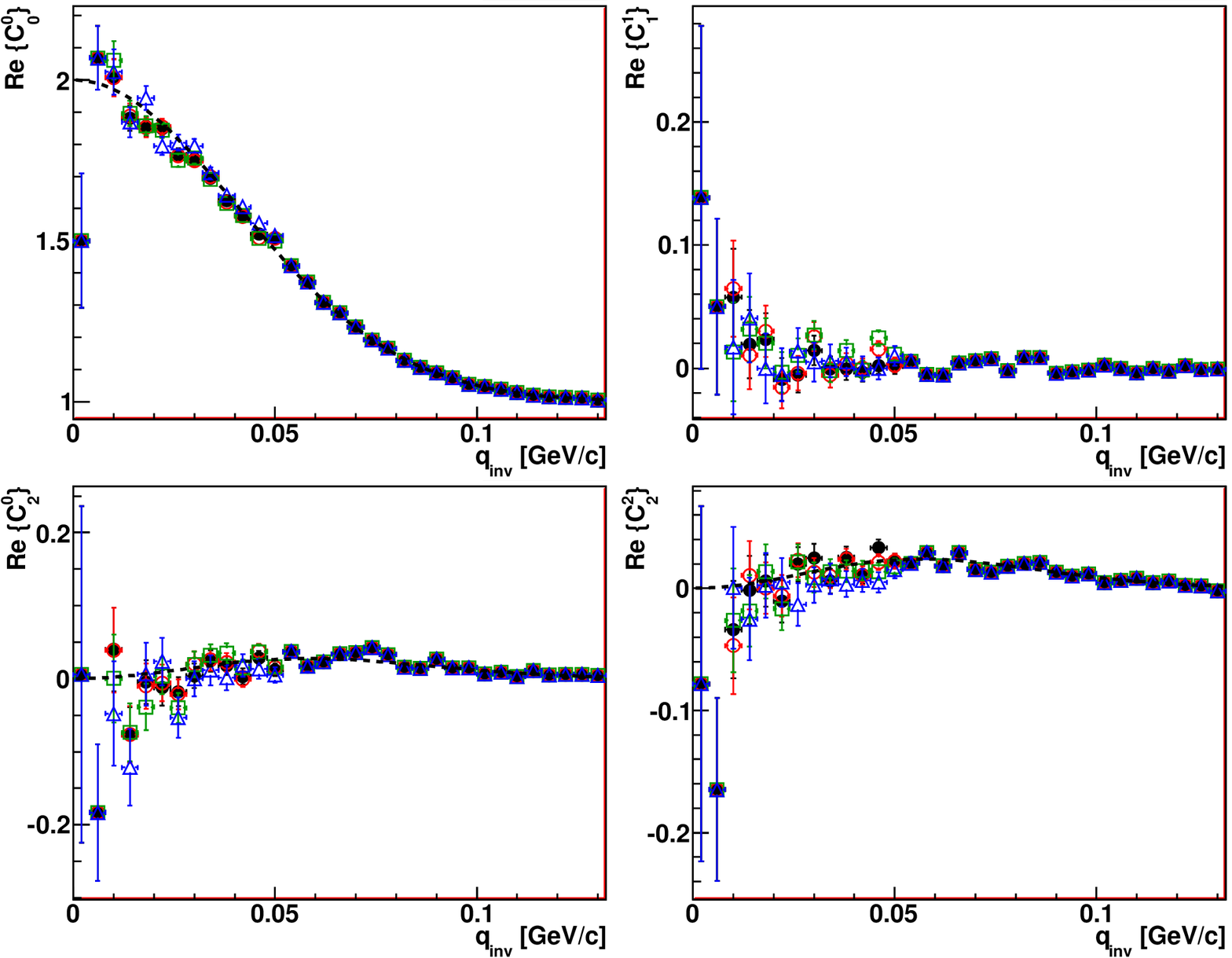}
%\vspace{-6.5mm}
\caption{(Color online) The simulation of the acceptance hole, made at
midrapidity (small $\cos\theta$) at small $q=2k^*$ (from $0.01$ to
$0.05$ GeV/c) and with varying width in $\phi$ (from $0$ to $3\pi/2$).
The plot shows the correlation in spherical harmonics for: 
ideal case with no hole (black diamonds), ``small hole'' $\pi/6$ (red circles), 
``sizeable hole'' $\pi/2$ (green squares), and ``huge hole'' $3\pi/2$ (blue
stars).  The black dashed lines show the analytical prediction for how the
spherical harmonics should look like for these sizes. }
\label{fig:cfhole}
\end{figure*}
\end{center}

To attempt to determine the practical limits of the technique,  
we have performed a test.  First, we calculate the
correlation function for identical pions using the Therminator model.  
We chose to use identical neutral pion pairs for the calculation simply because 
final state interactions do not distort the correlation appreciably, meaning that the 
correlation shape can be characterized simply by the correlation radii $R_{side}$, 
$R_{out}$ and $R_{long}$.  The source size has been set to reasonable 
values in the longitudinally co-moving system (of $\sim 3-4$ fm).  The 
first calculation does not have any acceptance holes.
We then repeat the calculation, introducing an artificial hole in the acceptance by removing both
from the numerator and denominator all pairs within the hole.  The
hole is at midrapidity (small $\cos\theta$), small $q$ (from $0.01$ to $0.05$
GeV/c) and with varying width in $\phi$ (from $0$ to $3\pi/2$). 
The results are shown in Fig.~\ref{fig:cfhole}. 

One can see that introducing the hole had no influence on the
extracted correlation function within statistical errors. Indeed, the
dominant effect of the acceptance hole has been to decrease
statistics, increasing statistical scatter and the corresponding
uncertainty. To further make this point, we show the analytical
prediction for how the spherical harmonics should look like for these
sizes  in the black dashed lines. As one can see, all points follow
the lines perfectly. 

Our results are in contrast to what would happen in the traditional approach of expanding
the correlation in spherical harmonics after making the ratio of 3D histograms.  
If there is poor statistics due to a gap in acceptance, then one will need 
a large number of spherical moments to capture the purely statistical 
fluctations present in the poorly populated high-$q$ bins.  What is more insidious, 
because the correlation is a ratio, the structure in poorly determined 3D bins 
appear to ``cancel out'' even when the poorly resolved data 
should not cancel out.  Rather, the poor statistics should give rise to large 
uncertainties and not contribute to the spherical harmonic expansion (which is what 
happens in our method).

\begin{center}
\begin{figure}[tb]
\includegraphics[angle=0,width=0.45 \textwidth]{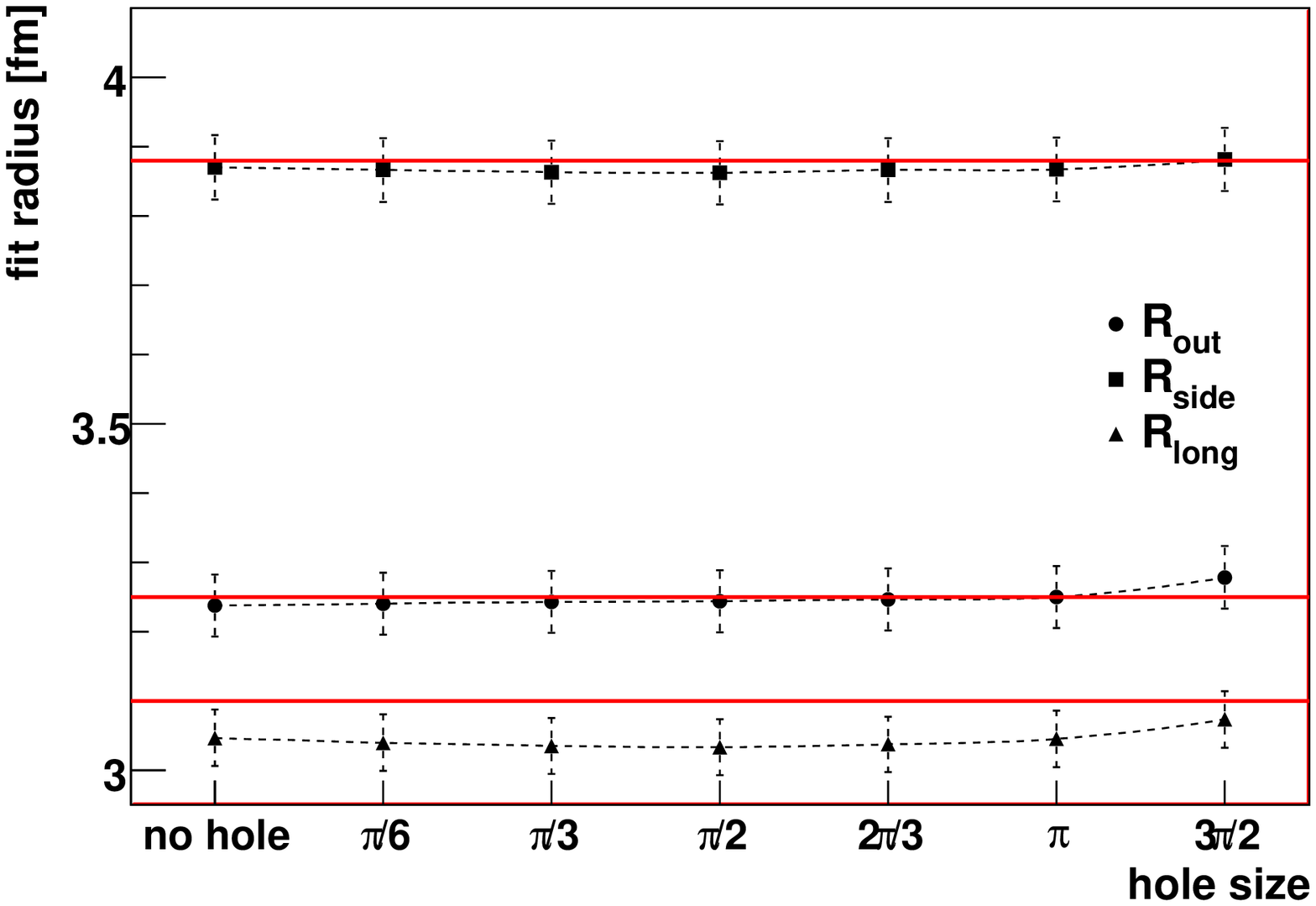}
%\vspace{-6.5mm}
\caption{(Color online) The values of the analytical fit to the
spherical harmonics vs. the hole size.  The solid lines are the ``input radii'' from 
the Therminator calculations and the dashed lines/symbols are our fits.}
\label{fig:cfrad}
\end{figure}
\end{center}

In Fig. \ref{fig:cfrad}, we show the values of the analytical fit to the
spherical harmonics vs. the hole size. The lines are the ``input radii'' from the
Therminator model. As one sees, the fit results are very stable and moreover in 
reasonable agreement with the input values.  Note: exact agreement between the input
and extracted radii can not be expected because the collective motion 
in the Therminator model shrinks the effective homogeneity length seen by 
the pairs and hence the correlation radii.  Even the very large hole of $3\pi/2$
(only a quarter of acceptance remaining!) our method seems to preserve all the
relevant components, provided that enough statistics remains outside the hole
region.

\subsection{Experimental corrections}

In order to be useful, the procedure for calculating the correlation
function directly in spherical harmonics should allow for the
application of the standard experimental corrections. Here we briefly
describe how this can be done.

Experimental resolution for two-particle reconstruction and
identification is usually dominated by two issues: track merging
(where two tracks in the detector are reconstructed as one) and track
splitting (where a single track is mistakenly reconstructed as
two).  These have non-trivial dependence on both the particle momenta
as well as their trajectory in the detector.  This is usually corrected
for by assigning a weight to each pair, based on the detailed detector
simulation.  Such weighting can be incorporated in the procedure in a
straightforward way.  When filling the numerator and the denominator
with pairs, one simply fills it with
the appropriate weight.  Mathematically it amounts to modifying
Eq.~\eqref{eq:tfrompairs} by multiplying the $Y^{*}_{lm}
(\Omega_{\hat q_i})$ by an additional weight $W$, coming from the
above mentioned correction.

Another common issue is the particle purity, namely the fraction of
pairs in the sample that should be treated as correlated.  A pair may
be not correlated if one of the particles is misidentified or if at
least one of the particles comes from a weak decay.  The experiment
should be able to estimate the the average purity of pairs $P$, which
can be (and usually is) a function of particles' momenta, and
therefore also of the pair relative momentum ${\bf q}$.  We 
use the traditional formula:
\begin{equation}
C_{corr}({\bf q}) = \frac{C_{meas}({\bf q}) - 1}{P({\bf q})} + 1.
\label{eq:purcor}
\end{equation}
From the correlation function $C$, we can obtain the correlation effect
$R \equiv C - 1$. In spherical harmonic representation this only
modifies the $\ell=0, m=0$ component: $R_{00} = C_{00} - 1$, while
others remain the same: $R_{\ell m} = C_{\ell m}$. Then
Eq.~\eqref{eq:purcor} simplifies to:
\begin{equation}
R_{corr}({\bf q}) = \frac {R_{meas}({\bf q})} {P({\bf q})}.
\label{eq:Rpurcor}
\end{equation}
We immediately note that it is equivalent to calculating the
correlation function from the numerator and denominator.  Therefore it
is enough to express purity $P$ directly in spherical harmonics and
treat it as denominator, take the measured correlation function and
treat it as numerator and finally apply the mathematical formalism
described in this work to obtain the correlation function corrected for
purity.

\section{Applicability}

The method presented in this paper has been successful applied to the
femtoscopic correlation functions in heavy-ion collisions. It should
be possible to apply it to other fields as well, however one has to
take into account limits of the method applicability. 

The basic formula~\eqref{eq:getClmDirectly} is strictly correct mathematically
only if one uses an infinite number of $\ell,m$ components for all the
functions ${\bf T}_q$, ${\bf M}_q$ and ${\bf C}_q$. In practical
application one needs to limit oneself to a specific value of $\ell$.
This is only allowed if the higher $\ell$-moments are negligible. The
femtoscopic correlation function is very well suited to the method
because the intrinsic symmetries of the pair distributions limit
the relevant $\ell$ components to a practical maximum of 6.  Most
important information is contained in $\ell=0$, $\ell=1$ and $\ell=2$
components. 

We have shown that the method remains stable for any reasonable acceptance hole 
in $\phi$ region.  It is also clear that the method will start breaking down 
only for really small values of $\phi-\theta$ acceptance and in the extreme case
of the ``hole'' taking up the whole $\phi-\theta$ acceptance the
method will simply break down due to lack of data.  The method in
effect interpolates the correlation function in the region where there
is no data by assuming certain symmetries in the underlying pair
distribution.  Using the method described here for femtoscopy, acceptance holes are 
``irrelevant'' - any reasonable femtoscopic measurement will have a large enough 
acceptance to be insensitive to the holes.

\section*{Acknowledgements}
The authors wish to thank Andrew Glenn for his careful reading of the manuscript.

This work performed under the auspices of the U.S. Department of Energy 
by Lawrence Livermore National Laboratory under Contract DE-AC52-07NA27344,
and by the U.S. NSF grant no. PHY-0653432.
\nobreak

\end{document}